\documentclass[runningheads]{llncs}
\usepackage[utf8]{inputenc}
\usepackage[T1]{fontenc}
\usepackage{graphicx}
\usepackage{grffile}
\usepackage{longtable}
\usepackage{wrapfig}
\usepackage{rotating}
\usepackage[normalem]{ulem}
\usepackage{amsmath}
\usepackage{textcomp}
\usepackage{amssymb}
\usepackage{capt-of}
\usepackage{hyperref}
\usepackage{listings}
\usepackage{tcolorbox}
\usepackage{subcaption}
\usepackage{libertine}
\usepackage{lastpage}
\usepackage{a4wide}
\usepackage{alltt}
 \PassOptionsToPackage{hyphens}{url}\usepackage{hyperref}
\newcommand{\swhurl}[1]{https://archive.softwareheritage.org/#1}
\newcommand{\swhref}[2]{\href{\swhurl{#1}}{#2}}

\definecolor{dkgreen}{rgb}{0,0.6,0}
\definecolor{gray}{rgb}{0.5,0.5,0.5}
\definecolor{mauve}{rgb}{0.58,0,0.82}
\author{Roberto Di Cosmo\inst{1}\orcidID{0000-0002-7493-5349}}  \authorrunning{R. Di Cosmo}
\institute{Software Heritage, Inria and University of Paris, France   \email{roberto@dicosmo.org}}
\date{March 2020}
\title{Archiving and referencing source code with Software Heritage}
\hypersetup{
 pdfauthor={dicosmo},
 pdftitle={Archiving and referencing source code with Software Heritage},
 pdfkeywords={},
 pdfsubject={},
 pdfcreator={Emacs 26.1 (Org mode 9.1.14)}, 
 pdflang={English}}
\begin{document}

\maketitle
\begin{abstract}
Software, and software source code in particular, is widely used in modern
research.  It must be properly archived, referenced, described and cited in
order to build a stable and long lasting corpus of scientific knowledge.  In
this article we show how the Software Heritage universal source code archive
provides a means to fully address the first two concerns, by archiving
seamlessly all publicly available software source code, and by providing
\emph{intrinsic persistent identifiers} that allow to reference it at various
granularities in a way that is at the same time convenient and effective.

We call upon the research community to adopt widely this approach.

\end{abstract}

\section{Introduction}
\label{sec:org051f9fd}

Software source code is \emph{an essential research output}, and there is a growing
general awareness of its importance for supporting the research process
\cite{Borgman2012,Stodden-reprod-2012,Hinsen2013}.  Many research communities
focus on the issue of \emph{scientific reproducibility} and strongly encourage
making the source code of the artefact available by archiving it in
publicly-accessible long-term archives; some have even put in place mechanisms
to assess research software, like the \emph{Artefact Evaluation} process introduced
in 2011 and now widely adopted by many computer science conferences
\cite{Dagstuhl-Artefacts-2016}, and the \emph{Artifact Review and Badging} program of
the ACM \cite{AcmBadges}. Other raise the complementary issues of making it
easier to discover existing research software, and giving academic credit to
authors \cite{force11citationprinciples,HowisonBullard2016,Lamprecht2019}.

As a first step, it is important to clearly identify the different concerns that
come into play when addressing software, and in particular its source code, as a
research output, that can be classified as follows:
\begin{quote}
\begin{description}
\item[{Archival:}] software artifacts must be properly \textbf{archived}, to ensure we can
\emph{retrieve} them at a later time;
\item[{Reference:}] software artifacts must be properly \textbf{referenced} to ensure we
can \emph{identify} the exact code, among many potentially archived
copies, used for reproducing a specific experiment;
\item[{Description:}] software artifacts must be equipped with proper \textbf{metadata} to
make it easy to \emph{find} them in a catalog or through a search
engine;
\item[{Citation:}] research software must be properly \textbf{cited} in research articles
in order to give \emph{credit} to the people that contributed to it.
\end{description}
\end{quote}

As already pointed out in the literature, these are not only different concerns,
but also \emph{separate} ones. Establishing proper \emph{credit} for contributors via
\emph{citations} or providing proper metadata to \emph{describe} the artifacts requires a
\emph{curation} process \cite{swmath,ASCL,dicosmo:hal-02475835} and is way more complex than
simply providing stable, intrinsic identifiers to \emph{reference} a precise version
of a software source code for reproducibility purposes
\cite{HowisonBullard2016,2020GtCitation,cise-2020-doi}.  Also, as remarked in
\cite{Hinsen2013,2020GtCitation}, resarch software is often a thin layer on top
of a large number of software dependencies that are developed and maintained
outside of academia, so the usual approach based on institutional archives is
not sufficient to cover all the software that is relevant for reproducibility of
research.\\

In this article, we focus on the first two concerns, \emph{archival} and \emph{reference},
showing how they can be adressed fully by leveraging the Software Heritage
universal archive \cite{swhcacm2018}, and also mention some recent evolutions in
best practices for embedding \emph{metadata} in software development repositories.

In Section \ref{sec:swh} we briefly recall what is Software Heritage and what
makes it special; in Section \ref{sec:archive} we show how researchers can
easily ensure that any relevant source code is archived; in Section
\ref{sec:reference} we explain how to use the \emph{intrinsic identifiers} provided
by Software Heritage to enrich research articles, making them more useful and
appealing for the readers, and providing stable links between articles and
source code in the web of scientific knowledge we are all building. Finally, we
point to ongoing collaborations and future perspectives in Section
\ref{sec:perspectives}.

\section{Software Heritage: the universal archive of software source code \label{sec:swh}}
\label{sec:orgeb67d36}

Software Heritage \cite{swhipres2017,swhcacm2018} is a non profit, long term
universal archive specifically designed for software source code, and able to
store not only a software artifact, but \emph{also its full development history}.\\

\begin{figure}[ht]
\centering
\includegraphics[width=.8\linewidth]{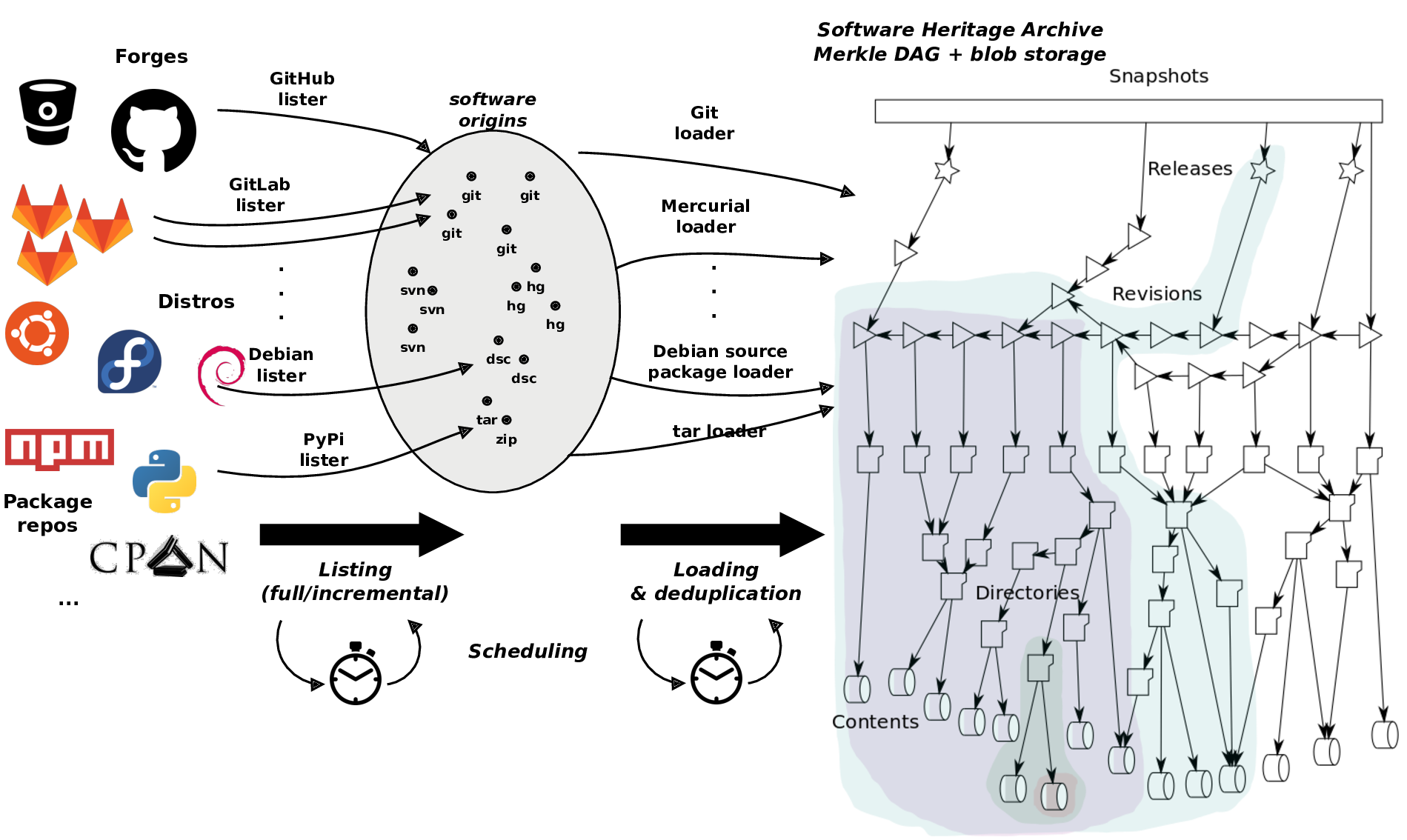}
\caption{\label{fig:orgcb7b0ab}
Architecture of the Software Heritage crawler}
\end{figure}

Software Heritage's mission is to collect, preserve, and make easily accessible
the source code of \emph{all} publicly available software. Among the strategies
designed for collecting the source code there is the development of a large
scale automated crawler for source code, whose architecture is shown in Figure
\ref{fig:orgcb7b0ab}.

We recall here a few key properties that set Software Heritage apart from all
other scholarly infrastructures:

\begin{itemize}
\item it \emph{proactively} archives \emph{all software}, making it possible to store and
reference any piece of publicly available software relevant to a research
result, independently from any specific field of endeavour, and even when
the author(s) did not take any step to have it archived \cite{swhipres2017,swhcacm2018};\\

\item it stores the source code with its development history in a uniform
data structure, a Merkle DAG \cite{Merkle}, that allows to provide
uniform, \emph{intrinsic} identifiers for the billions of software artifacts
of the archive, independently of the version control system or package
format used \cite{cise-2020-doi}.
\end{itemize}

At the time of writing this article, the Software Heritage archive contains
over 7 billions unique source code files, from more than 100 million different
software 
origins\footnote{See \url{https://archive.softwareheritage.org} for the up to date figures.}.
It provides the ideal place to \emph{preserve research
software artifacts}, and offers powerful mechanisms to \emph{enhance research
articles} with precise references to relevant fragments of your source code.
Using Software Heritage is straightforward and involves very simple steps, 
that we detail in the following sections.

\section{Archiving and self archiving \label{sec:archive}}
\label{sec:orge0e2234}

In a research article one may want to reference different kinds of source code
artifacts: some may be popular open source components, some may be general
purpose libraries developed by others, and some may be one own's software
projects.\\

All these different kinds of software artifacts can be archived
extremely easily in Software Heritage: it's enough that their source code is
hosted on a publicly accessible repository (Github, Bitbucket, any GitLab
instance, an institutional software forge, etc.) using one of the version
control systems supported by Software Heritage, currently Subversion, Mercurial
and Git
\footnote{For up to date information, see \url{https://archive.softwareheritage.org/browse/origin/save/}}.\\

\begin{figure}[h!]
\begin{center}
\includegraphics[width=\textwidth]{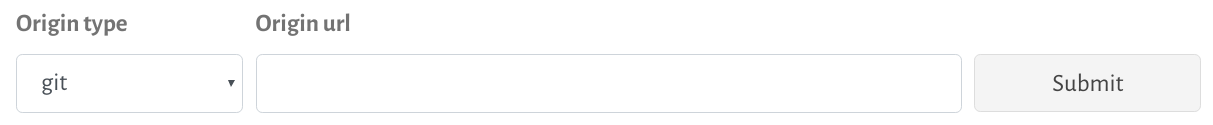}
\end{center}
\caption{The Save Code Now form}\label{fig:savecodenow}
\end{figure}

For source code developed on popular development platforms, chances are that
the code one wants to reference is already archived in Software Heritage, but
one can make sure that the archived version history is fully up to date, as
follows:

\begin{itemize}
\item go to \url{https://save.softwareheritage.org},
\item pick the right version control system in the drop-down list, enter the code repository url
\footnote{Make sure to use the clone/checkout url as given by the development platform hosting your code. It can easily be found in the web interface of the development platform.},
\item click on the Submit button (see Figure \ref{fig:savecodenow}).
\end{itemize}

\noindent That's all. No need to create an account or disclose personal
information of any kind. If the provided URL is correct, Software Heritage will
archive the repository shortly after, with its full development history.  If it
is hosted on one of the major forges we already know, this process will take
just a few hours; if it is in a location we never saw before, it can take
longer, as it will need to be manually
screened \footnote{It is also possible to request archival programmatically, using the
Software Heritage API, which can be quite handy to integrate in a Makefile;
see \url{https://archive.softwareheritage.org/api/1/origin/save/} for details.}.

\subsection{Preparing source code for self archiving}
\label{sec:org1873ed0}
\label{ssec:prepare} In case the source code is one own's, before requesting
its archival it is important to structure the software repository follow well
established good practices for release management \cite{raymond2013}. In
particular one should add \texttt{README} and \texttt{AUTHORS} files as well as
licence information following industry standard terminology \cite{reuse,SPDXLicences}.

Future users that find the artifact useful might want to give credit by citing
it.  To this end, one might want to provide instructions on how one prefers the
artifact to be cited. We would recommend to also provide structured metadata
information in machine readable formats. While practices in this area are still
evolving, one can use the CodeMeta generator available at
\url{https://codemeta.github.io/codemeta-generator/} to produces metadata
conformant to the CodeMeta schema: the JSON-LD output can be put at the root of the
project in a \textbf{codemeta.json} file. Another option is to use the Citation File Format,
CFF (usually in a file named \textbf{citation.cff}).

\section{Referencing \label{sec:reference}}
\label{sec:orgb9710f8}
\begin{figure}[htbp]
\centering
\includegraphics[width=.8\linewidth]{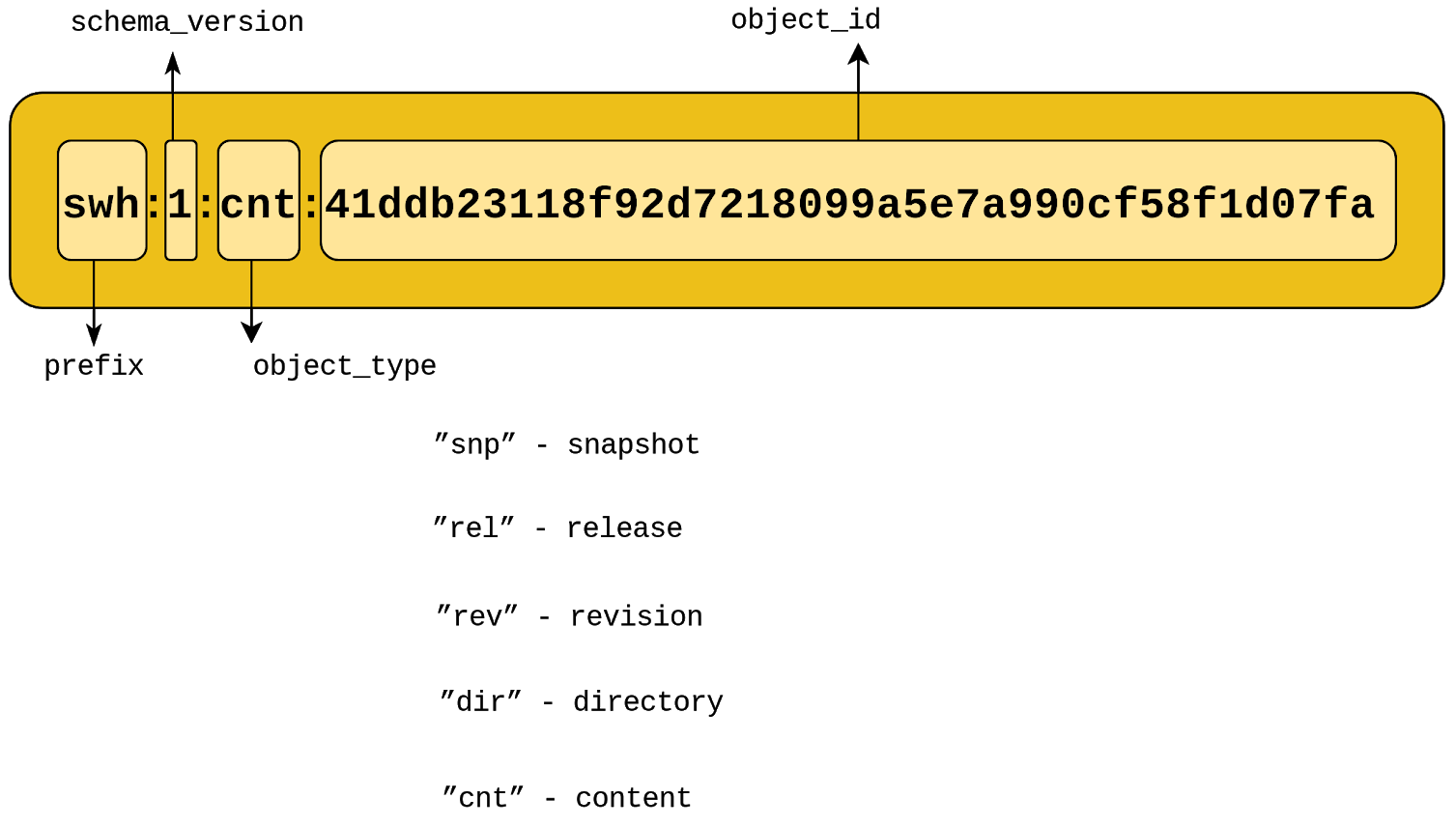}
\caption{\label{fig:orgf31a4db}
Schema of the core Software Heritage identifiers}
\end{figure}

Once the source code has been archived, the Software Heritage \emph{intrinsic
identifiers}, called SWH-ID, \href{https://docs.softwareheritage.org/devel/swh-model/persistent-identifiers.html}{fully documented online} and shown in Figure
\ref{fig:orgf31a4db}, can be used to reference with great ease any version of
it.

SWH-IDs are URIs with a very simple schema: the \texttt{swh} prefix makes
 explicit that these identifiers are related to Software Heritage; the colon
 (\verb|:|) is used as separator between the logical parts of identifiers; the
 schema version (currently \verb|1|) is the current version of this identifier
 schema; then follows the type of the objects identified and finally comes a
 hex-encoded (using lowercase ASCII characters) cryptographic signature of this
 object, computed in a standard way, as detailed in \cite{swhipres2018,cise-2020-doi}.

These core identifiers may be equipped with the \emph{qualifiers} that carry
contextual \emph{extrinsic} information about the object:

\begin{description}
\item[{origin :}] the \emph{software origin} where an object has been found or observed in the wild, as an URI;\\
\item[{visit :}] persistent identifier of a \emph{snapshot} corresponding to a specific \emph{visit} of a repository containing the designated object;\\
\item[{anchor :}] a \emph{designated node} in the Merkle DAG relative to which a \emph{path to the object} is specified;\\
\item[{path :}] the \emph{absolute file path}, from the \emph{root directory} associated to the \emph{anchor node}, to the object;\\
\item[{lines :}] \emph{line number(s)} of interest, usually within a content object
\end{description}

The combination of the core SWH-IDs with these qualifiers provides a very
powerful means of referring in a research article to all the software artefacts
of interest.\\

To make this concrete, in what follows we use as a running example the article
\emph{A ``minimal disruption'' skeleton experiment: seamless map and reduce
embedding in OCaml} by Marco Danelutto and Roberto Di Cosmo \cite{Parmap2012}
published in 2012. This article introduced an elegant library for multicore
parallel programming that was distributed via the \url{gitorious.org}
collaborative development platform, at \url{gitorious.org/parmap}. Since
Gitorious has been shut down a few years ago, like Google Code and CodePlex,
this example is particularly fit to show why pointing to an \emph{archive} of the
code is better than pointing to the collaborative development platform where it
is developed.

\subsection{Specific version}
\label{sec:org86ce8f5}

The Parmap article describes a \emph{specific version} of the Parmap library, the one
that was used for the experiments reported in the article, so in order to
support reproducibility of these results, we need to be able to pinpoint
precisely the state(s) of the source code used in the article.

The exact revision of the source code of the library used in the article
has the following SWH-ID:

\begin{tcolorbox}
swh:1:rev:0064fbd0ad69de205ea6ec6999f3d3895e9442c2;\\
origin=https://gitorious.org/parmap/parmap.git;\\
visit=swh:1:snp:78209702559384ee1b5586df13eca84a5123aa82
\end{tcolorbox}

This identifier can be turned into a clickable URL by prepending to it the
prefix \url{https://archive.softwareheritage.org/} (one can try it by clicking on \href{https://archive.softwareheritage.org/swh:1:rev:0064fbd0ad69de205ea6ec6999f3d3895e9442c2;origin=https://gitorious.org/parmap/parmap.git;visit=swh:1:snp:78209702559384ee1b5586df13eca84a5123aa82}{this link}).
\subsection{Code fragment}
\label{sec:orgb0970c7}

\begin{figure}[t]
\centering
    \begin{subfigure}[t]{0.49\textwidth}
        \centering
        \includegraphics[scale=0.5]{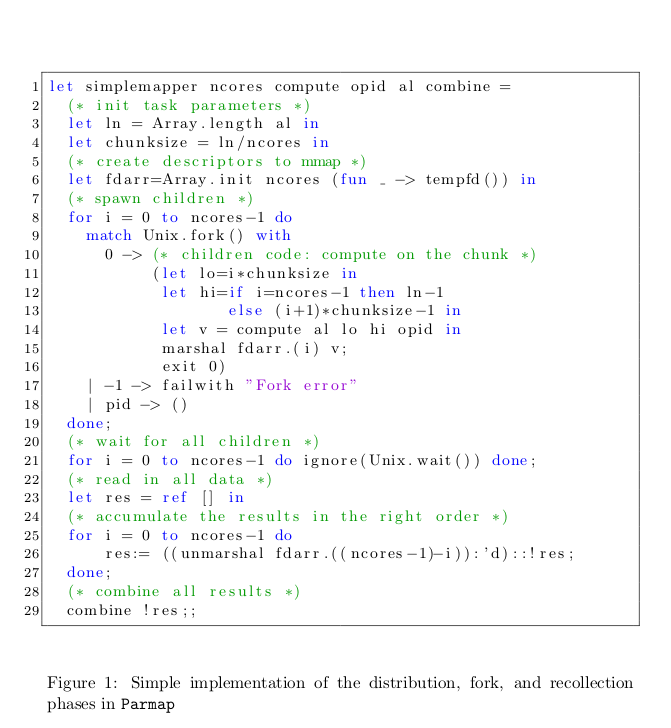}
        \caption{as presented in the article \cite{Parmap2012}}\label{fig:parmappaper}
    \end{subfigure}
    \vspace{1em}
    \begin{subfigure}[t]{0.49\textwidth}
        \centering
        \includegraphics[scale=0.48]{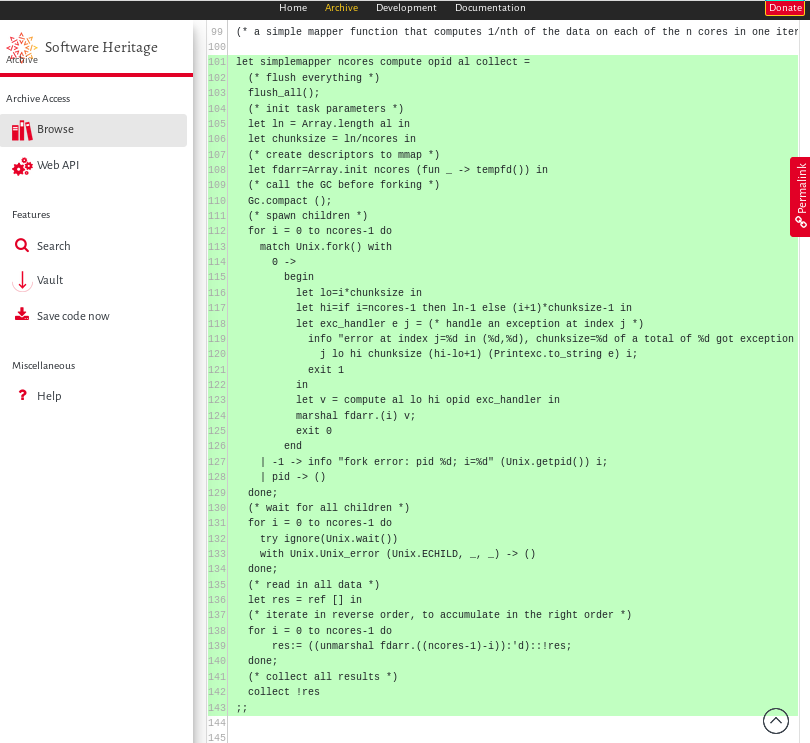}
        \caption{as archived in Software Heritage}\label{fig:parmapswh}
    \end{subfigure}
    \caption{Code fragment from the published article compared to the content
      in the Software Heritage archive}
  \label{fig:parmap}
\end{figure}

Having a link to the exact archived revision of a software project is
important in all research articles that use software, and the core SWH-IDs
allow to drill down and point to a given directory or even a file content,
but sometimes, like in our running example, one would like to do more, and
pinpoint a fragment of code inside a specific version of a file. This is
possible using the \texttt{lines=} qualifier available for identifiers that
point to file content.

Let's see this feature at work in our running example, showing how the
experience of studying or reviewing an article can be greatly enhanced by
providing pointers to code fragments.

In Figure 1 of \cite{Parmap2012}, which is shown here as Figure
\ref{fig:parmappaper}, the authors want to present the core part of the code
implementing the parallel functionality that constitutes the main contribution
of their article. The usual approach is to typeset in the article itself \emph{an
excerpt of the source code}, and let the reader try to find it by delving
into the code repository, which may have evolved in the mean time.
Finding the exact matching code can be quite difficult, as the code excerpt
is \emph{often edited} a bit with respect to the original, sometimes to drop
details that are not relevant for the discussion, and sometimes due to
space limitations.\\

In our case, the article presented 29 lines of code, slightly edited from the
43 actual lines of code in the Parmap library: looking at
\ref{fig:parmappaper}, one can easily see that some lines have been dropped
(102-103, 118-121), one line has been split (117) and several lines simplified
(127, 132-133, 137-142).\\

Using Software Heritage, the authors can do a much better job, because the original
code fragment can now be precisely identified by the following Software Heritage identifier:
\begin{tcolorbox}
swh:1:cnt:d5214ff9562a1fe78db51944506ba48c20de3379;\\
origin=https://gitorious.org/parmap/parmap.git;\\
visit=swh:1:snp:78209702559384ee1b5586df13eca84a5123aa82;\\
anchor=swh:1:rev:0064fbd0ad69de205ea6ec6999f3d3895e9442c2;\\
path=/parmap.ml;\\
lines=101-143
\end{tcolorbox}
This identifier will \textbf{always} point to the code fragment shown in Figure \ref{fig:parmapswh}.\\

The caption of the original article shown in Figure \ref{fig:parmappaper} can
then be significantly enhanced by incorporating a clickable link containing
the SWH-ID shown above: it's all is needed to point to the exact source code
fragment that has been edited for inclusion in the article, as shown in Figure
\ref{fig:swhall}. The link contains, thanks to the SWH-ID qualifiers, all
the contextual information necessary to identify the context in which this
code fragment is intended to be seen.
\begin{figure}[h!]
\begin{tcolorbox}
Simple implementation of the distribution, fork, and recollection phases in \texttt{Parmap} 
(slightly simplified from the \href{https://archive.softwareheritage.org/swh:1:cnt:d5214ff9562a1fe78db51944506ba48c20de3379;origin=https://gitorious.org/parmap/parmap.git;visit=swh:1:snp:78209702559384ee1b5586df13eca84a5123aa82;anchor=swh:1:rev:0064fbd0ad69de205ea6ec6999f3d3895e9442c2;path=/parmap.ml;lines=101-143}{the actual code in the version of Parmap used for this article})
\end{tcolorbox}
\caption{A caption text with the link to the code fragment and its contextual information}\label{fig:swhall}
\end{figure}

When clicking on the hyperlinked text in the caption shown above, the reader is
brought seamlessly to the Software Heritage archive on a page showing the
corresponding source code archived in Software Heritage, with the relevant
lines highlighted (see Figure \ref{fig:parmapswh}).\\

\subsection{Getting the SWH-ID}
\label{sec:org2f3349d}
\label{ssec:getswhid} 
\begin{figure}[hbt]
  \centering
  \includegraphics[width=\linewidth]{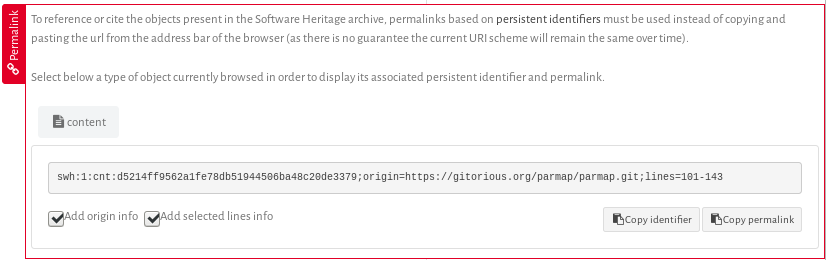}
  \caption{Obtaining a Software Heritage identifier using the permalink box on
    the archive Web user interface}
  \label{fig:permalink}
\end{figure}

A fully qualified SWH-ID is rather long, and it needs to
 be, as it contains quite a lot of information that is essential to convey.  In
 order to make it easy to use SWH-IDs, we provide a very simple way of getting
 the right SWH-ID without having to type it by hand. Just browse the archived
 code in Software Heritage and navigate to the software artifact of
 interest. Clicking on the \emph{permalinks vertical red tab} that is present on all
 pages of the archive, opens up a tab that allows to select the identifier for
 the object of interest: an example is shown in Figure \ref{fig:permalink}.

The two buttons on the botton right allow to copy the identifier or the full permalink
in the clipboard, and to paste it in an article as needed.

\subsection{Generating and verifying SWH-IDs}
\label{sec:org7d2599e}

An important consequence of the fact that SWH-IDs are \emph{intrinsic identifiers} is
that they can be generated and verified \emph{independently} of Software Heritage,
using \texttt{swh-identify}, an open source tool developed by Software Heritage,
and distributed via PyPI as \texttt{swh.model}, with the stable version at the
time of writing being
\swhref{swh:1:rev:919db4f50e500d1c63879ddab20cf4fc1346c275;origin=https://pypi.org/project/swh.model/;visit=swh:1:snp:9f7a900d46f609f60c194d142abef1965ff28e02}\{this
one\}.\\

Version 1 of the SWH-IDs uses git-compatible hashes, so if the source code that
one wants to reference uses git as a version control system, one can create the
right SWH-ID by just prepending \texttt{swh:1:rev:} to the commit hash. This
comes handy to automate the generation of the identifiers to be included in an
article, as one will always have code and article in sync. 

\section{Perspectives for the scholarly world \label{sec:perspectives}}
\label{sec:orge5a673e}

We have shown how Software Heritage and the associated SWH-IDs enables the
seamless archival of all publicly available source code. It provides for all
kind of software artifacts the \emph{intrinsic identifiers} that are needed to
establish long lasting, resilient links between research articles and the
software they use or describe.\\

All researchers can use \emph{right now} the mechanisms presented here to produce
improved and enhanced research articles. More can be achieved by establishing
collaborations with academic journals, registries and institutional repositories
and registries, in particular in terms of description and support for software
citation. Among the intial collaborations that have been already established, we
are happy to mention the cross linking with the curated mathematical software
descriptions maintained by the \texttt{swMath.org} portal \cite{swmath}, and the
curated deposit of software artefacts into the HAL french national open access
portal \cite{dicosmo:hal-02475835}, which is performed via a standard SWORD
protocol inteface, an approach that is currently being explored by other
academic journals.

We believe that the time has come to see software become a first class citizen
in the scholarly world, and Software Heritage provides a unique infrastructure
to support an open, non profit, long term and resilient web of scientific
knowledge.

\subsection{Acknowledgements}
\label{sec:org4eb2112}
This article is a major evolution of the research software archival and
reference guidelines available on the Software Heritage website
\cite{swh-archive-guide} resulting from extensive discussions that took place
over several years with many people. Special thanks to Alain Girault, Morane
Gruenpeter, Antoine Lambert, Julia Lawall, Arnaud Legrand, Nicolas Rougier and
Stefano Zacchiroli for their precious feedback on these issues and/or earlier
versions of this document.

\clearpage
\bibliography{biblio}
\bibliographystyle{abbrv}

\appendix\clearpage
\end{document}